\documentclass[pra,superscriptaddress,twocolumn]{revtex4}

\usepackage{enumerate}
\usepackage{amsfonts,amssymb,amsmath}
\usepackage[]{graphics,graphicx,epsfig}
\usepackage{amsthm}

\usepackage{graphicx}
\usepackage{dcolumn}
\usepackage{natbib}
\usepackage{color}
\usepackage{multirow}
\usepackage{ulem}

\begin{document}


\title{Attractors and asymptotic dynamics of open discrete-time quantum walks on cycles}

\author{A. S. Sajna}
\affiliation{Faculty of Physics, Adam Mickiewicz University, Umultowska 85, 61-614 Pozna\'n, Poland}
\affiliation{Department of Physics, Boston University, 590 Commonwealth Ave., Boston, MA 02215, USA} 

\author{T. P. Polak}
\affiliation{Faculty of Physics, Adam Mickiewicz University, Umultowska 85, 61-614 Pozna\'n, Poland} 

\author{A. W{\'o}jcik} 
\affiliation{Faculty of Physics, Adam Mickiewicz University, Umultowska 85, 61-614 Pozna\'n, Poland}

\author{P. Kurzy{\'n}ski}
\email{pawel.kurzynski@amu.edu.pl}
\affiliation{Faculty of Physics, Adam Mickiewicz University, Umultowska 85, 61-614 Pozna\'n, Poland}

\date{\today}


\begin{abstract}
Open quantum walks often lead to a classical asymptotic behavior. Here, we look for a simple open quantum walk whose asymptotic behavior can be non-classical. We consider a discrete-time quantum walk on n-cycle subject to a random coin-dependent phase shift at a single position. This finite system, whose evolution is described by only two Kraus operators, can exhibit all kinds of asymptotic behavior observable in quantum Markov chains: it either evolves towards a maximally mixed state, or partially mixed state, or tends to an oscillatory motion on an asymptotic orbit. We find that the asymptotic orbits do not have a product structure, therefore the corresponding states can manifest entanglement between the position and the coin degrees of freedom, even if the system started in a product state.  
\end{abstract}

\maketitle


\section{Introduction}

Evolution of a closed quantum system is governed by the Hamiltonian $H$. The state of the system evolves according to $e^{-i H \Delta t}|\psi(t)\rangle \equiv U(\Delta t)|\psi(t)\rangle=|\psi(t+\Delta t)\rangle$, where $U(\Delta t)$ is the unitary evolution operator and $\Delta t$ is the time of the evolution. The Hamiltonian is often considered to be primary and the unitary evolution operator to be secondary, in a sense that $H$ determines $U$, not the other way around. Therefore, the concepts like energy or momentum seem to be more fundamental in physics than a change in time or a translation in space. 

However, change and translation are much more intuitive. The way a system changes in time or travels through space can be defined via a simple rule. This allows to formulate dynamics in an algorithmic way, i.e., describe it via a set of instructions that need to be applied to a system to evolve it forward in time. One of the most prominent applications of this approach in classical physics is known as cellular automata (CA). CA provide a universal dynamical framework that can be used to model various physical phenomena \cite{CA,CA2}.  

There were attempts to quantize classical CA \cite{QCA0}, but as far as we know discrete-time quantum walks (DTQWs) \cite{Aharonov,QCA} are considered to be the most successful of them. DTQWs are simple models describing dynamics of a single particle in a discrete space. There is an intuitive unitary rule which translates the particle to a neighbouring position. The direction of the translation is determined by a state of an auxiliary system that is known as a coin. Due to the underlying quantum nature, the particle and the coin can be in a superposition and as a result the particle spreads in all directions in a coherent way. The model appeared to be very powerful and found applications in various fields of quantum physics and quantum information science. An interested reader can learn more about it from a collection of review papers \cite{Review1,Review2,Review3}.

DTQWs evolve according to unitary rules, which are fundamentally reversible. As a consequence, for small finite systems one cannot observe an emergence of any stable complex structures. This is because reversible dynamics is quasiperiodic. It always comes back close to the initial state. This is a consequence of the Poincar\'e recurrence theorem. Of course, the Poincar\'e recurrence time scales with the size of the system and for very large systems one may observe interesting emergent behaviors before the recurrence happens. In particular, it was shown that if the DTQW takes place in the infinite space one can observe that the coin degree of freedom tends to a stationary state \cite{ThermQW}. Nevertheless, for DTQWs defined in few-dimensional Hilbert spaces one can observe recurrences if the evolution is traced for sufficiently long times. 

Fortunately, the DTQW unitary rules can be complemented with irreversible ones \cite{OpenQW1,OpenQW2,OpenQW3,OpenQW4,OpenQW5}. The first irreversible rules introduced to DTQWs were modelling the effect of decoherence \cite{DecohQW1,DecohQW2,DecohRev}. However, although finite space DTQWs with decoherence tend to stationary states, these states are the same as the ones of the corresponding classical random walks. It is therefore natural to look for irreversible processes leading to new quantum behaviors. 

Our main motivation is to look for a simple modification of standard unitary DTQW rules that would give rise to attractors and non-classical asymptotic behavior. We focus on DTQWs on n-cycles, i.e., a one-dimensional discrete space with $n$ positions and periodic boundary conditions. We supplement the standard unitary evolution with only two Kraus operators. The first one is proportional to identity and the second one is a position and coin-dependent phase shift -- its action is nontrivial only if the particle is at a certain location and the coin is in a certain state. Such Kraus operators are in fact unitary transformations that act with some probability, therefore the process is described by a random unitary channel. 

Random unitary channels were already considered in the DTQW model, but in the context of dynamically percolated graphs \cite{PercolationQW1,PercolationQW2,PercolationQW3,PercolationQW4}.  These works descent from more general studies on asymptotic dynamics generated by random unitary channels \cite{RU1,RU2}, or even more general quantum Markov chains \cite{QMC1,QMC2}. Such irreversible processes were shown to either converge to a stationary state, or tend to persistent oscillations on an attractive orbit. Moreover, it was shown that random unitary dynamics can generate entanglement (see for example \cite{RUEnt}), therefore it is possible to observe a non-classical asymptotic behavior in open quantum dynamics. 

We find that, although our model is described by three parameters, it is enough to manipulate only one of them to observe all of the aforementioned asymptotic behaviors of quantum Markov chains. More precisely, we observe: either i) convergence to a maximally mixed stationary state,  or ii) convergence to a partially mixed stationary state, or iii) persistent oscillations on an attractive orbit. In the last case, the attractive orbit lies in a subspace that does not admit a decomposition into coin and position subspaces, hence the system initiated in a product state can eventually fall onto attractive orbit made of entangled states. 


\section{The model}

We consider a discrete-time quantum evolution of a particle whose dynamics is governed by a two-level system, known as a coin. The coin can be an intrinsic degree of freedom of the particle (alike to spin-1/2 or polarization). The particle moves in one-dimensional discrete space with periodic boundary conditions and its state is given by $|x\rangle$ ($x=1,2,\ldots,n$). The state of the coin is $|c\rangle$ ($c=0,1$) and the joint system is described by $|x\rangle\otimes|c\rangle$. 

A single step of the evolution is determined by a unitary operator 
\begin{equation}\label{U}
U=S(\openone_x \otimes C), 
\end{equation}
which is a coin rotation $C$ chosen here to be
\begin{eqnarray}
C|0\rangle &=& \frac{1}{\sqrt{2}} (|0\rangle - |1\rangle), \\
C|1\rangle &=& \frac{1}{\sqrt{2}} (|0\rangle + |1\rangle),
\end{eqnarray} 
followed by the conditional translation $S$
\begin{eqnarray}
S|x\rangle\otimes|0\rangle &=& |x+1\rangle\otimes|0\rangle, \\
S|x\rangle\otimes|1\rangle &=& |x-1\rangle\otimes|1\rangle.
\end{eqnarray} 
Periodic boundary conditions imply $|n+1\rangle \equiv |1\rangle$. In the above $\openone_x$ is the identity operator on the position space. The state after $T$ steps is given by
\begin{equation}
|\psi_T\rangle = U^{T}|\psi_0\rangle.
\end{equation}
The above evolution generates entanglement between the coin and the position. This is because of the conditional translation operator
\begin{equation}
S|x\rangle\otimes(\alpha |0\rangle + \beta|1\rangle) = \alpha |x+1\rangle\otimes|0\rangle + \beta|x-1\rangle\otimes|1\rangle. 
\end{equation}   

Next, consider the following Kraus operators 
\begin{eqnarray}
K_0 &=& \sqrt{1-\eta} \openone_x \otimes \openone_c, \label{K0} \\
K_1 &=& \sqrt{\eta} \left[ (\openone_x - |n\rangle\langle n|) \otimes \openone_c \right. \nonumber \\
&+& \left. |n\rangle\langle n| \otimes (e^{i\varphi_0}|0\rangle\langle 0| + e^{i\varphi_1}|1\rangle \langle 1|)\right], \label{K1}
\end{eqnarray}
where $\openone_c$ is the identity on the coin space. These Kraus operators are described by three parameters: $\eta\in [0,1]$ and $\varphi_0 ,\varphi_1 \in [0,2\pi)$. For $\eta=0$ the Kraus operators have no effect. For $\eta=1$ they unitarily apply the coin-dependent phase shifts $\varphi_0$ and $\varphi_1$, if particle is at position $x=n$. In any other case the phase shifts are applied with probability $\eta$. Note, that $K_0^{\dagger}K_0 + K_1^{\dagger}K_1 = \openone_x\otimes \openone_c$.

Together with the unitary operation (\ref{U}), the Kraus operators (\ref{K0}) and (\ref{K1}) generate the following irreversible evolution
\begin{equation}\label{evolution}
\rho(t+1) = K_0 U \rho(t) U^{\dagger} K_0^{\dagger} + K_1 U \rho(t) U^{\dagger} K_1^{\dagger},
\end{equation}
where $\rho(t)$ is the density matrix of the system at time $t$. 

Note, that if the operator $K_1$ were position independent and of the form
\begin{equation}
K_1 = \sqrt{\eta} \openone_x \otimes \sigma_z, 
\end{equation}
where $\sigma_z$ is Pauli-Z operator, the system would undergo standard decoherence causing the transition of the DTQW into a classical random walk \cite{DecohRev}. In particular, if $\eta = 1/2$ the full decoherence would happen in a single step and the corresponding DTQW would be equivalent to a classical random walk on n-cycle. In this model the particle moves randomly, either one step to the right or left and each possibility occurs with probability $1/2$. For such process the system tends to a stationary state that is uniformly distributed over all positions and over both coin states $\frac{1}{2n}\openone_x \otimes \openone_c$. We are going to show that our model with a position and coin-dependent phase shift can arrive at this stationary state. However, we are also going to show that there is a specific set of parameters $\varphi_0$ and $\varphi_1$ for which the system has different asymptotic behavior.

Before we proceed, let us remark that for even cycles the system effectively performs a walk on only half of positions. This is because in this case probability amplitudes at even positions never interfere with the ones at odd positions, therefore the evolution can be separated into two different walks. That is why we focus only on odd $n$.


\section{Dynamics}

In this section we present our preliminary observations obtained from numerical simulations of system's dynamics. A detailed analysis is going to be presented in the following section. We found that the asymptotic behavior of the system can be divided into two general cases, depending on the choice of parameters $\varphi_0$ and $\varphi_1$. Moreover, we found that one of the parameters, say $\varphi_0$, can be fixed and it is enough to manipulate with $\varphi_1$ to observe all kinds of asymptotic behavior.


\subsection{The case $\varphi_0 \neq 0$ and $\varphi_1 \neq 0$}

\begin{figure}[t]
\includegraphics[scale=0.30]{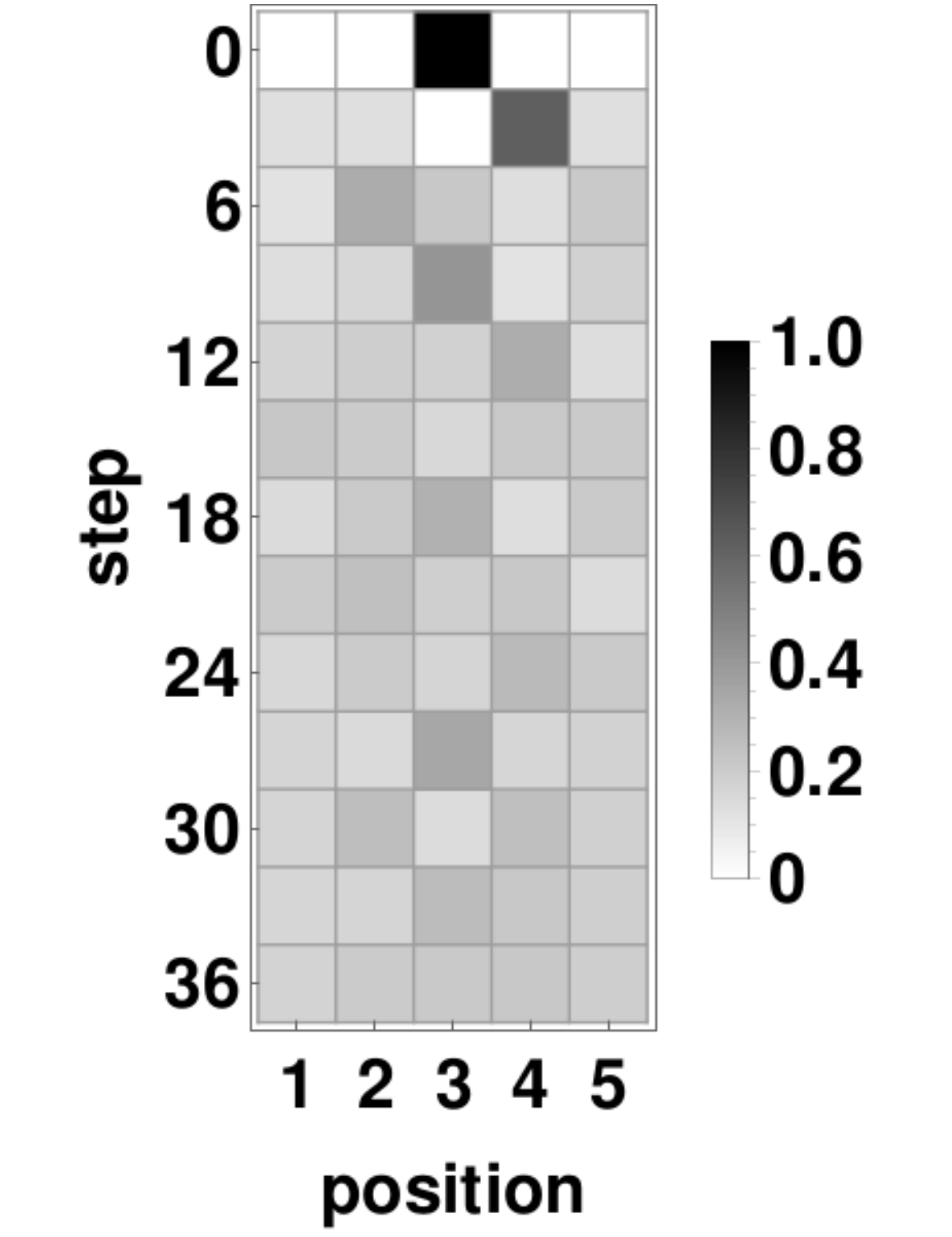}\includegraphics[scale=0.35]{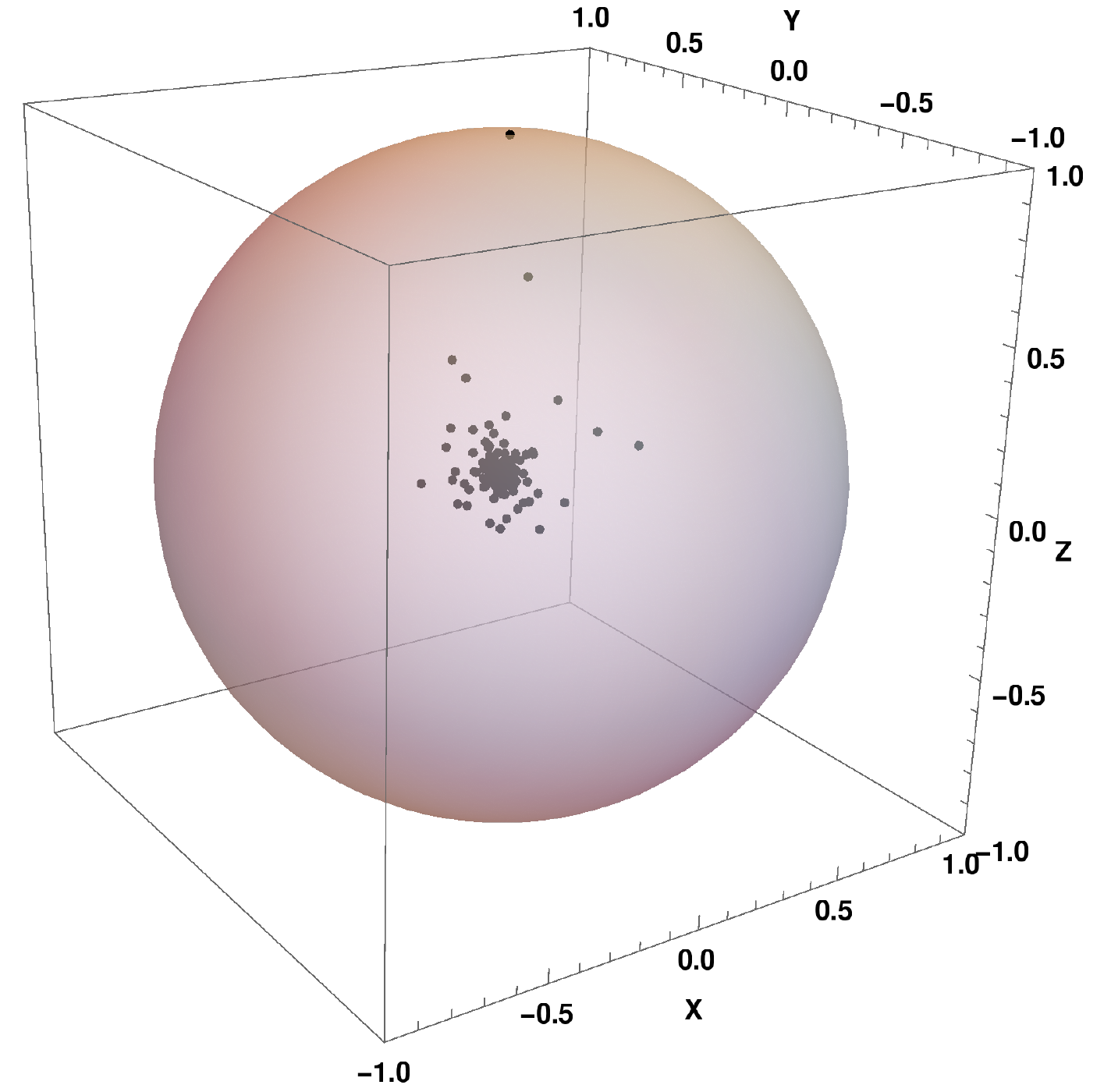}
\caption{The evolution of the DTQW on 5-cycle for $\eta = 1/2$, $\varphi_0=\pi/2$, $\varphi_1=\pi/3$ and the initial state $|3\rangle\otimes|0\rangle$. The system tends to a maximally mixed state $\frac{1}{10}\openone_x\otimes\openone_c$. The left plot represents the evolution of the spatial probability distribution for the first 36 steps (showing every third step). The right plot shows the Bloch ball representation of the evolution of the coin degree of freedom. Dots inside the ball correspond to the coin states for the first 100 steps and one can see that they start to gather in the centre. \label{fig1}}
\end{figure}

We observed that if both phase shifts are non-zero the system tends to a stationary state. The parameters $\eta$, $\varphi_0$ and $\varphi_1$ determine the mixing time -- the closer $\varphi_{0/1}$ to $\pi$ and $\eta$ to $1/2$ the faster the relaxation to a stationary state happens. In addition, the form of the stationary state is determined by $\varphi_0$ and $\varphi_1$. For $\varphi_0 \neq \varphi_1$ the systems relaxes to a maximally mixed state $\rho(\infty)=\frac{1}{2n}\openone_x\otimes\openone_c$. Therefore, for a wide range of parameters the system tends to a classical behavior. An example of such behavior for 5-cycle is presented in Fig. \ref{fig1}.

\subsection{The case $\varphi_0 = \varphi_1 \neq 0$}

For $\varphi_0 = \varphi_1 \neq 0$ the stationary state may not be maximally mixed. Its form depends on the initial state. In particular, the non-maximally mixed stationary state can be observed if the system is initialized at the position $x=n$, i.e., the position at which the phase shifts are applied. In this case the reduced state of the position is $\frac{1}{n}\openone_x$ and the stationary state of the coin $\rho_c$ depends on the initial coin state $\rho_0$. We found the following fitting  
\begin{eqnarray}
\rho_0 &=& \begin{pmatrix}
\frac{1+\cos\theta}{2} & \gamma\frac{\sin\theta}{2}e^{i\alpha} \\ \gamma\frac{\sin\theta}{2}e^{-i\alpha} & \frac{1-\cos\theta}{2}
\end{pmatrix} \rightarrow \nonumber \\
\rho_c &=& \begin{pmatrix}
\frac{1}{2} & i \gamma\frac{\sin\theta}{2n}\sin\alpha \\ -i \gamma\frac{\sin\theta}{2n}\sin\alpha & \frac{1}{2}
\end{pmatrix},\label{fitting}
\end{eqnarray}
where $\gamma \in [0,1]$ determines the purity of the initial state. In the next section we will provide analytical arguments for this behavior.


\subsection{The case $\varphi_0 \neq 0$ and $\varphi_1 = 0$}

\begin{figure}[t]
\includegraphics[scale=0.30]{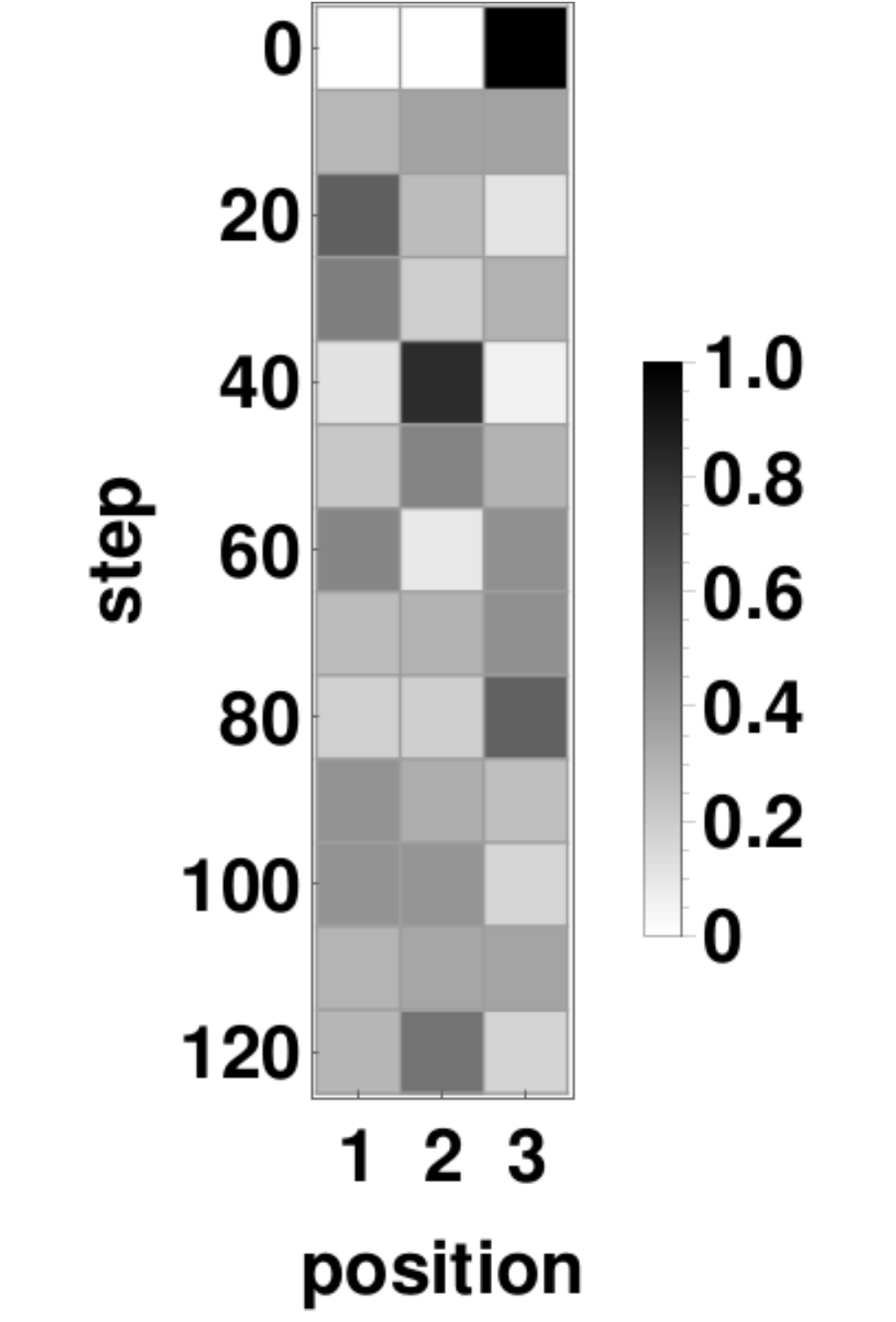}\includegraphics[scale=0.35]{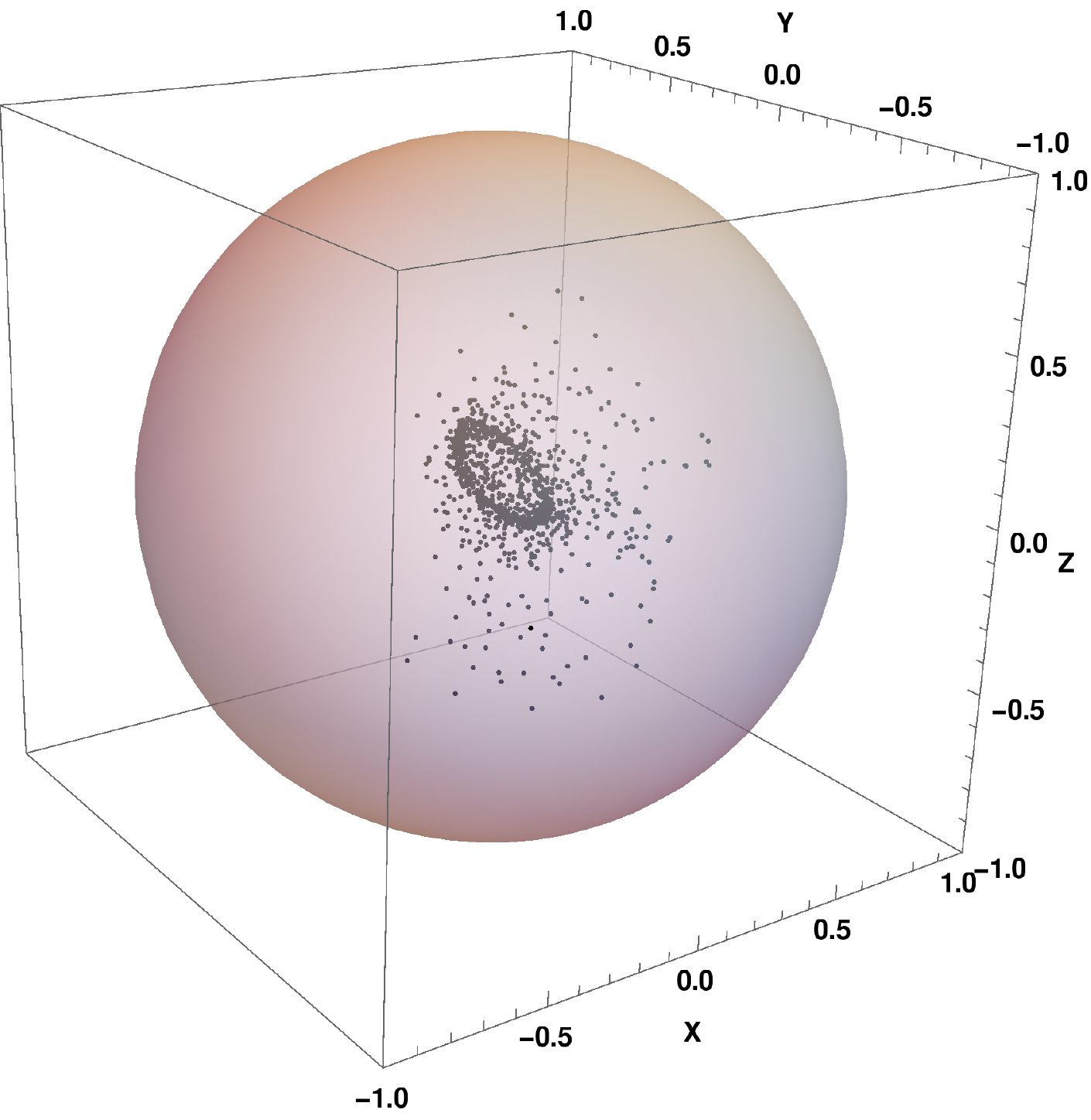}
\caption{The evolution of the DTQW on 3-cycle for $\eta = 1/2$, $\varphi_0=\pi/10$, $\varphi_1=0$ and the initial state $|3\rangle\otimes\frac{1}{\sqrt{2}}(|0\rangle + e^{-i\pi/3}|1\rangle)$. The system does not tend to any stationary distribution. Instead, after short period of mixing it arrives at some quasi-periodic evolution. The left plot represents the evolution of the spatial probability distribution for the first 120 steps (showing every tenth step). The right plot shows the Bloch ball representation of the evolution of the coin degree of freedom. Dots inside the ball correspond to the coin states for the first 1000 steps and one can see the ellipsoidal attractor pattern that starts to emerge near the centre of the Bloch ball.\label{fig2}}
\end{figure}

The other type of behavior occurs for  $\varphi_1 = 0$ (or due to coin-flip symmetry for $\varphi_0=0$). In this case the system does not tend to a stationary state. Assuming that it starts in a pure state, it first looses its purity and gets mixed, but not maximally, and then it continues to evolve in a seemingly reversible way. This evolution occurs in both, position and coin space, i.e., the spatial probability distribution and the Bloch vector of the coin do not freeze, but continue to follow a quasi-periodic sequence of states on a closed orbit (see Fig. \ref{fig2}). Therefore, the system tends to an attractor which looks like a limit cycle. 

\begin{figure}[t]
\includegraphics[scale=0.3]{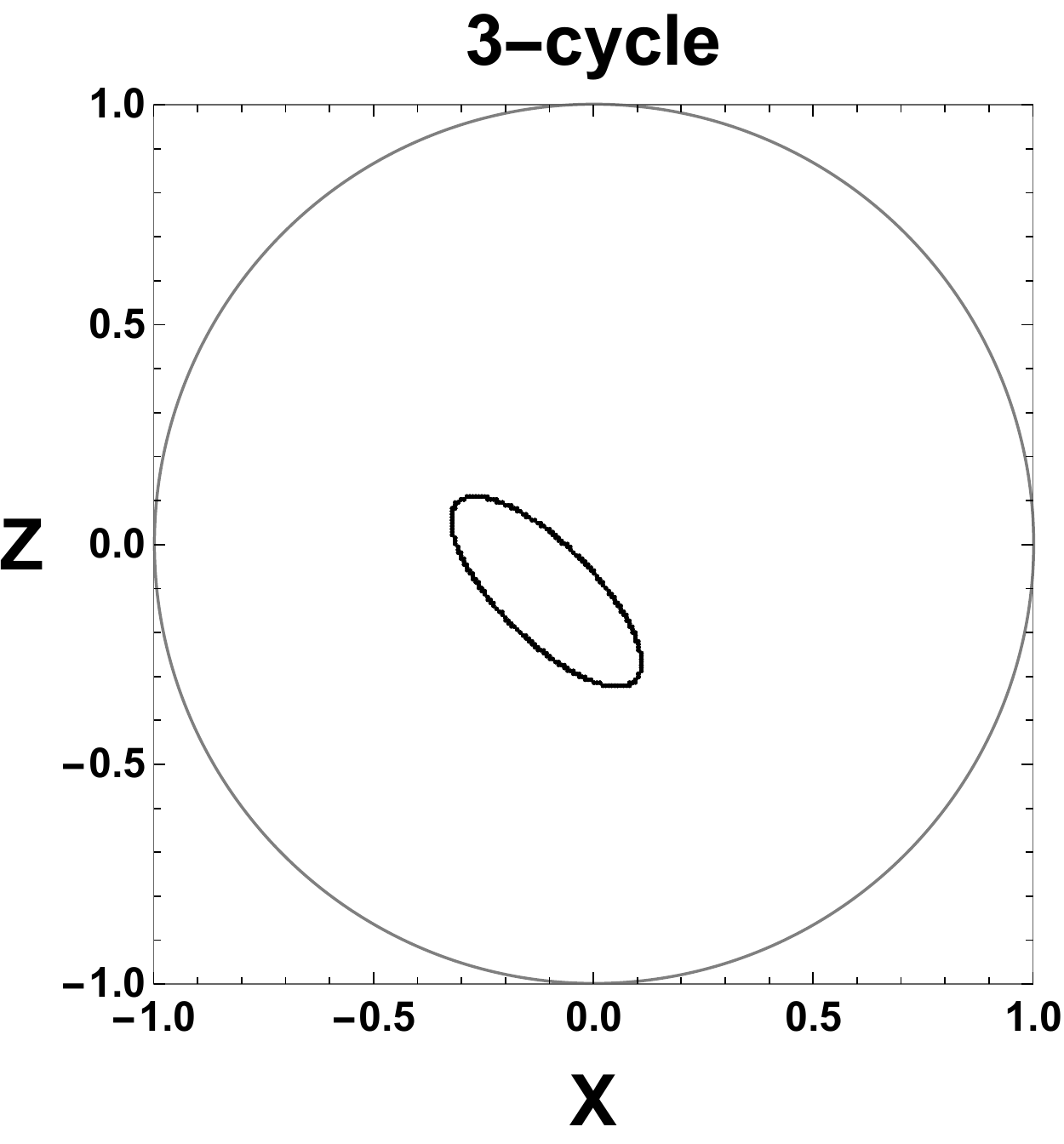}~~\includegraphics[scale=0.3]{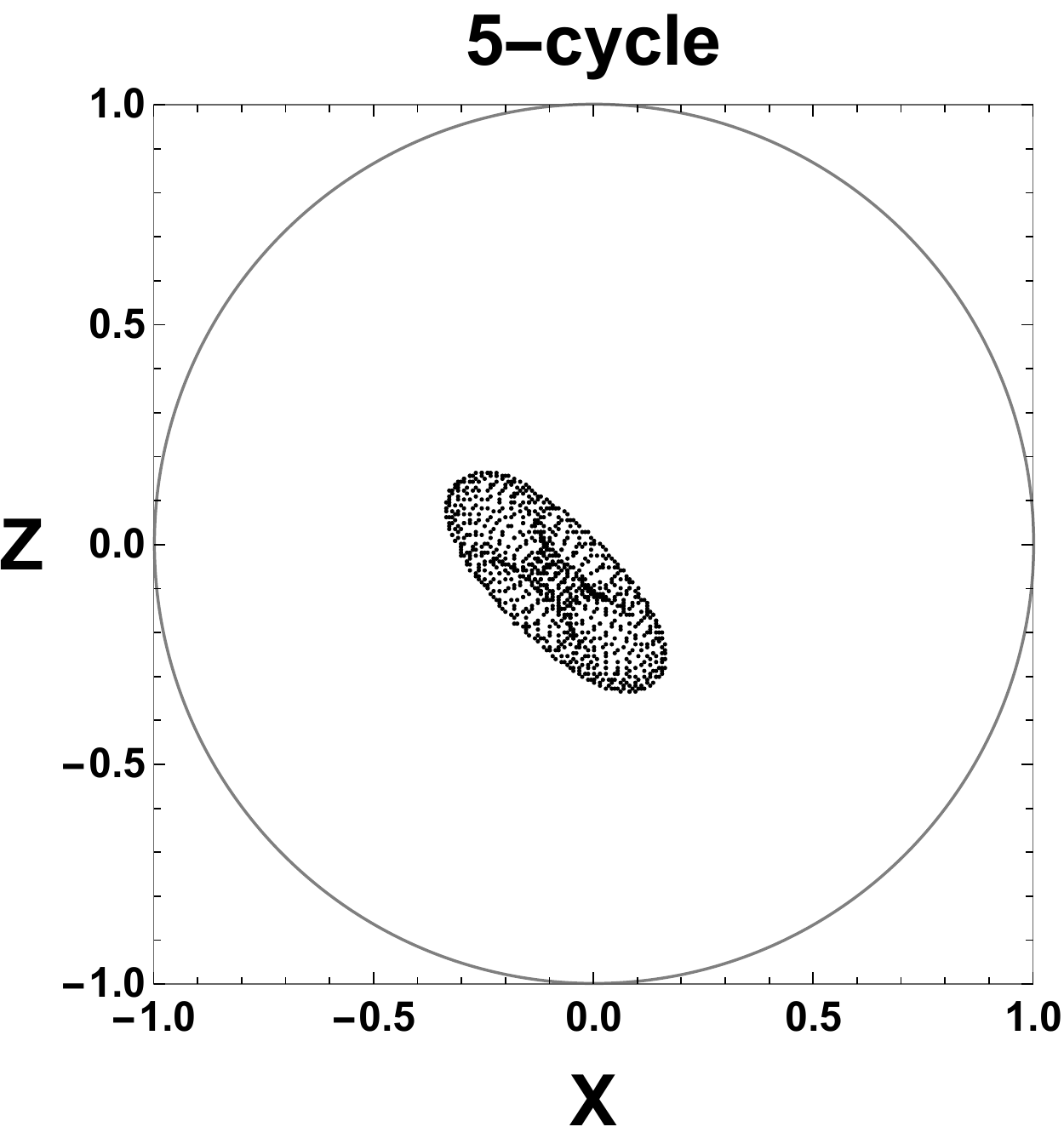}
\center{\includegraphics[scale=0.3]{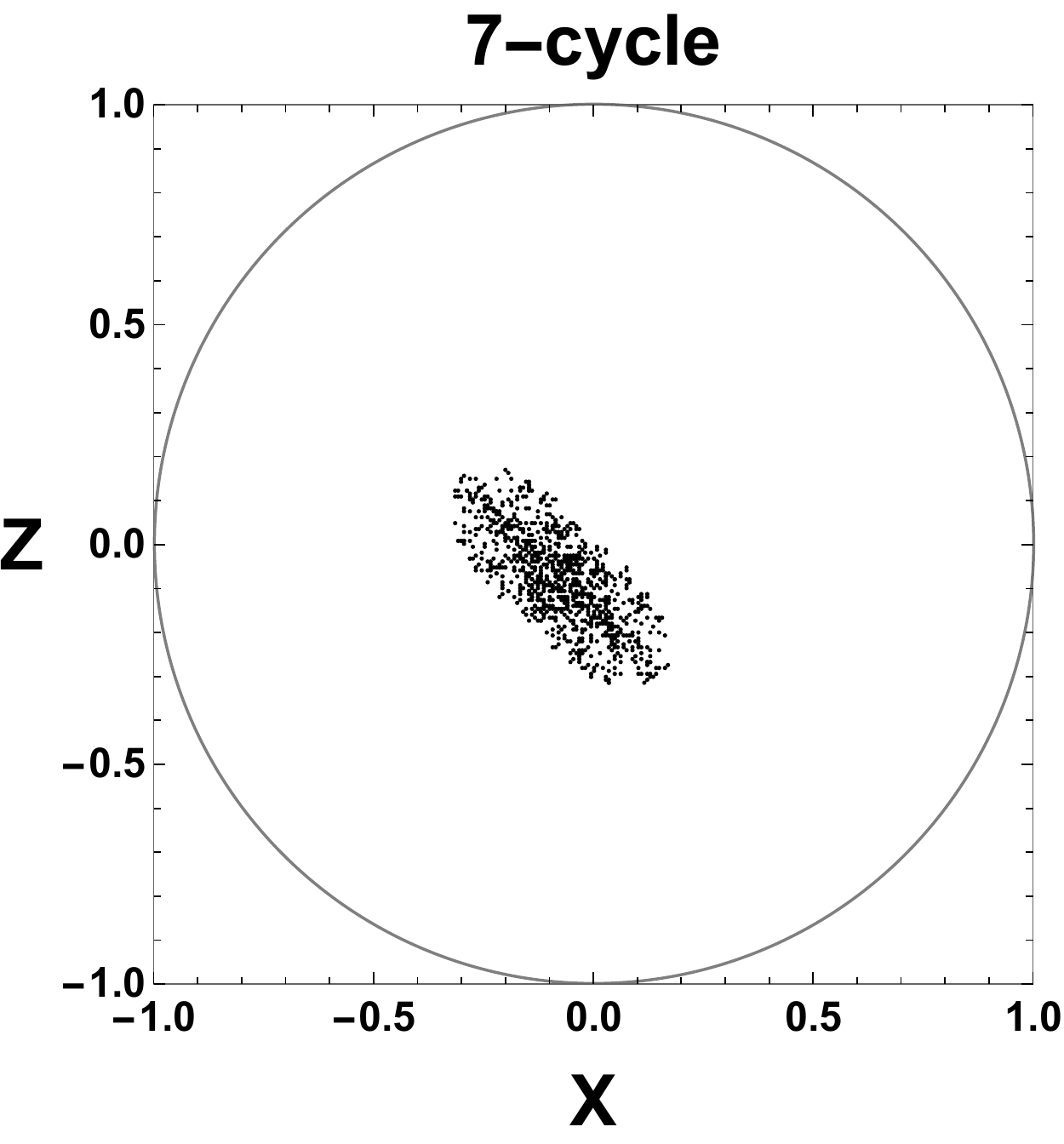}}
\caption{The XZ-section of the Bloch ball showing the evolution of the coin degree of freedom from $T=1000$ till $T=2000$. Dots correspond to subsequent coin states. The parameters are $\eta = 1/2$, $\varphi_0=\pi/2$, $\varphi_1=0$ and the initial state is $|1\rangle\otimes\frac{1}{\sqrt{2}}(|0\rangle + e^{-i\pi/3}|1\rangle)$. Top left: in case of 3-cycle the Bloch vector follows the pseudo-reversible evolution on an ellipsoid. Top right and bottom: in case of 5-cycle and 7-cycle the Bloch vector follows the pseudo-reversible evolution on a more complicated structure. \label{fig3}}
\end{figure}

The Bloch ball representation allows to visualize the reduced asymptotic dynamic on the attractor for the coin degree of freedom. It strongly depends on the size of the n-cycle on which the DTQW takes place. For  example, it can be an ellipsoid (3-cycle), a set of points forming a pattern (5-cycle), or a seemingly structureless set of points that is nevertheless confined to a finite region (7-cycle). These are depicted in Fig. \ref{fig3}.  

\section{Analysis}

In this section we provide an explanation of the above behaviors by analyzing the properties of the evolution generated by (\ref{evolution}). In particular, we show that the properties of the model result from the interplay between the spectral decomposition of the unitary evolution operator (\ref{U}) and the parameters $\varphi_0$ and $\varphi_1$.


\subsection{Random unitary channel}

Let us first observe that
\begin{eqnarray}
K_0 U & \equiv & \sqrt{1-\eta} U_0, \\
K_1 U & \equiv & \sqrt{\eta} U_1, \label{U1}
\end{eqnarray}
where $U_0=U$ and $U_1=VU$ are both unitary operators. The first one represents the standard DTQW evolution and the second one represents the DTQW evolution followed by a unitary coin-dependent phase shift  
\begin{eqnarray}
V &=& (\openone_x - |n\rangle\langle n|) \otimes \openone_c  \nonumber \\ &+&  |n\rangle\langle n| \otimes (e^{i\varphi_0}|0\rangle\langle 0| + e^{i\varphi_1}|1\rangle \langle 1|). \label{V}
\end{eqnarray}
A single step of the evolution is therefore given by
\begin{equation}\label{ruc}
\rho(t+1) = (1-\eta)U_0\rho(t)U_0^{\dagger} + \eta U_1\rho(t)U_1^{\dagger},
\end{equation}
i.e., the state evolves according to a randomly chosen unitary operator, either $U_0$ or $U_1$. 

The stationary states of the above evolution have special properties. Assume that $\sigma$ is a stationary state 
\begin{equation}
\sigma = (1-\eta) U_0 \sigma U_0^{\dagger}  + \eta U_1 \sigma U_1^{\dagger}.
\end{equation}
In this case the following also holds \cite{RU1,RU2}
\begin{equation}
\sigma = U_0 \sigma U_0^{\dagger} = U_1 \sigma U_1^{\dagger}.
\end{equation}
The above provides a substantial simplification of the asymptotic dynamics problem. In particular, we look for operators $X_{\lambda}$ that are eigenvectors of both transformations
\begin{equation}\label{oev}
U_0 X_{\lambda} U_0^{\dagger} = U_1 X_{\lambda} U_1^{\dagger} = \lambda X_{\lambda},
\end{equation}
where $\lambda$ is the eigenvalue. For random unitary channels these eigenvalues obey $|\lambda| \leq 1$ \cite{RU1,RU2}. For many steps $(t \rightarrow \infty)$ the following holds
\begin{equation}\label{asympt}
\rho_{\infty}(t) \approx \sum_{\lambda} \alpha_{X_{\lambda}} \lambda^t X_{\lambda},
\end{equation}  
where 
\begin{equation}
\alpha_{X_{\lambda}} = \text{Tr} \left(X_{\lambda}^{\dagger} \rho(0)\right).
\end{equation}

In the above we assumed that the operators $X_{\lambda}$ are normalized to one. It is clear that for $|\lambda| < 1$ and for large $t$ one observes $\lambda^t \rightarrow 0$, hence the sum in (\ref{asympt}) should take into account only operators $X_{\lambda}$ for which $|\lambda | = 1$. This is a further simplification, since we do not need to look for all eigenvectors. Moreover, Eq. (\ref{asympt}) clearly shows that the asymptotic evolution is unitary \cite{RU1,RU2}. The set of operators $\{X_{\lambda}\}_{|\lambda|=1}$ gives rise to an attractor of the dynamics. Such dynamics is oscillatory, however if all of these operators correspond to $\lambda = 1$, the attractor is a fixed point \cite{RU1,RU2}.  

Plugging $U_0=U$ and $U_1=VU$ to Eq. (\ref{oev}) leads to
\begin{equation}\label{evu}
UX_{\lambda} U^{\dagger} = \lambda X_{\lambda} 
\end{equation}  
and
\begin{equation}\label{evv}
VX_{\lambda} V^{\dagger} = X_{\lambda}.
\end{equation}
We provide solutions to these equations in the following subsections.


\subsection{Attractor space and asymptotic dynamics for $\varphi_0 \neq  0$ and $\varphi_1\neq 0$}

Due to Eqs. (\ref{V}) and (\ref{evv}) any operator $X_{\lambda}$ should have a particular block form that depends on the choice of $\varphi_0$ and $\varphi_1$. For $\varphi_0 \neq  0$, $\varphi_1\neq 0$ and $\varphi_0 \neq \varphi_1$ the operator $V$ has eigenvalues: $e^{i\varphi_0}$, $e^{i\varphi_1}$ and 1. The last eigenvalue is $2(n-1)$ times degenerate, therefore the block form of $X_{\lambda}$ is $2(n-1)\oplus 1 \oplus 1$, where the one-dimensional spaces are spanned by $|n\rangle \otimes |0\rangle$ and $|n\rangle \otimes |1\rangle$, respectively. Such block form implies that 
\begin{equation}\label{con1}
(\langle x |\otimes \langle c| ) X_{\lambda} (|n\rangle \otimes |c'\rangle) = (\langle n |\otimes \langle c'| ) X_{\lambda} (|x\rangle \otimes |c\rangle) = 0.
\end{equation}
for $x=1,\ldots,n-1$ and $c,c'=0,1$. In addition, the following should also hold 
\begin{equation}\label{con2}
(\langle n |\otimes \langle 0| ) X_{\lambda} (|n\rangle \otimes |1\rangle) = (\langle n |\otimes \langle 1| ) X_{\lambda} (|n\rangle \otimes |0\rangle) = 0.
\end{equation}

For $\varphi_0 = \varphi_1\neq 0$ the eigenvalues of $V$ are $e^{i\varphi_0}$ (two times degenerate) and $1$ ($2(n-1)$ times degenerate) and the block form of  $X_{\lambda}$ is $2(n-1)\oplus 2$, where the two-dimensional space is spanned by the same vectors as in the previous case. This time the operator $X_{\lambda}$ needs to obey (\ref{con1}), but not (\ref{con2}).

In the most general form we have
\begin{equation}
X_{\lambda} = \sum_{x,x'=1}^n \sum_{c,c'=0}^1 \gamma_{x,x',c,c'} |x\rangle \langle x'|\otimes |c\rangle\langle c'|.
\end{equation}
However, due to (\ref{con1}) $\gamma_{n,x,c,c'}=\gamma_{x,n,c,c'}=0$ for $x =1,\ldots,n-1$ and arbitrary $c$ and $c'$. Let us plug the above equation into (\ref{evu}). It is convenient to introduce coin states $|\pm \rangle = \frac{1}{\sqrt{2}}(|1\rangle \pm |0\rangle)$ and to occasionally use labels $c,c'=\pm$. Note, that $U$ transforms $|x-1\rangle \langle x'-1|\otimes |+\rangle\langle +|$ into $|x\rangle \langle x'|\otimes |0\rangle\langle 0|$. In a similar way, $|x-1\rangle \langle x'+1|\otimes |+\rangle\langle -|$, $|x+1\rangle \langle x'-1|\otimes |-\rangle\langle +|$, and $|x+1\rangle \langle x'+1|\otimes |-\rangle\langle -|$ are transformed into $|x\rangle \langle x'|\otimes |0\rangle\langle 1|$, $|x\rangle \langle x'|\otimes |1\rangle\langle 0|$, and $|x\rangle \langle x'|\otimes |1\rangle\langle 1|$, respectively. Therefore, (\ref{con1}) is satisfied iff
\begin{equation}\label{cn1}
\gamma_{x,n-1,c,+} = \gamma_{x,1,c,-} = \gamma_{n-1,x,+,c} = \gamma_{1,x,-,c} = 0,
\end{equation}
for $x=2,\ldots, n-2$ and arbitrary $c$. But in addition, the constraint (\ref{con1}) implies that after the application of $U$ the following holds
\begin{equation}\label{cn2}
\gamma_{x,n-1,c,1} = \gamma_{x,1,c,0} = \gamma_{n-1,x,1,c} = \gamma_{1,x,0,c} = 0,
\end{equation}
for $x=2,\ldots, n-2$ and arbitrary $c$. Both, (\ref{cn1}) and (\ref{cn2}), imply
\begin{equation}
\gamma_{x,n-1,c,c'} = \gamma_{x,1,c,c'} = \gamma_{n-1,x,c,c'} = \gamma_{1,x,c,c'} = 0,
\end{equation}
for $x=2,\ldots, n-2$ and arbitrary $c$ and $c'$.

One can follow the same procedure to show that
\begin{equation}
\gamma_{x,n-2,c,c'} = \gamma_{x,2,c,c'} = \gamma_{n-2,x,c,c'} = \gamma_{2,x,c,c'} = 0,
\end{equation}
for $x=3,\ldots, n-3$ and arbitrary $c$ and $c'$. Eventually, one can show  
\begin{equation}
\gamma_{x,n-k,c,c'} = \gamma_{x,k,c,c'} = \gamma_{n-k,x,c,c'} = \gamma_{k,x,c,c'} = 0,
\end{equation}
for $x=k+1,\ldots, n-k-1$ and arbitrary $c$ and $c'$. Therefore, the only operators that satisfy (\ref{con1}) are either
\begin{equation}
X_{\lambda}^{(1)} = \sum_{x=1}^n \sum_{c,c'=0}^1 \gamma_{x,c,c'}^{(1)} |x\rangle \langle x| \otimes |c\rangle\langle c'|,
\end{equation} 
or
\begin{equation}
X_{\lambda}^{(2)} = \sum_{x=1}^n \sum_{c,c'=0}^1 \gamma_{x,c,c'}^{(2)} |x\rangle \langle n-x| \otimes |c\rangle\langle c'|,
\end{equation}
or a linear combination of the two. It is straightforward to show that
\begin{eqnarray}
X_{\lambda=1}^{(1)} &=&  \openone_x \otimes \openone_c, \label{X1} \\
X_{\lambda=1}^{(2)} &=&  \sum_{x=1}^n |x\rangle \langle n-x| \otimes \sigma_y, \label{X2}
\end{eqnarray}
where $\sigma_y$ is Pauli-Y operator, and both operators are hermitian and correspond to the eigenvalue $\lambda=1$. We skipped the normalization coefficients, so that the norm ${\text Tr}\{X^{\dagger}X\}$ of both operators is $2n$.

Finally, note that while both operators satisfy the constraint (\ref{con1}), the constraint (\ref{con2}) is satisfied only by $X_{\lambda=1}^{(1)}$. Therefore, for $\varphi_0 \neq  0$, $\varphi_1\neq 0$ and $\varphi_0 \neq \varphi_1$ the attractor space consists of only one operator, which is proportional to identity on the whole space. As a result, the asymptotic dynamics in this case is just a fixed point corresponding to a maximally mixed state. This explains the behavior observed during the numerical simulations. 

On the other hand, for $\varphi_0 \neq  0$, $\varphi_1\neq 0$ and $\varphi_0 = \varphi_1$ the asymptotic dynamics is also a fixed point, but this time its form depends on the initial state $\rho_0$
\begin{equation}\label{asymptotic}
\rho_{\infty} = \frac{1}{2n} (X_{\lambda=1}^{(1)} + \xi X_{\lambda=1}^{(2)}),
\end{equation}
where $\xi = {\text Tr}\{X_{\lambda=1}^{(2)} \rho_0\}$. This also confirms our observations drawn from numerical simulations. In particular, the corresponding asymptotic coin state can have non-zero Bloch vector that points in the Y direction if it started at position $x=n$. The asymptotic state is a separable mixture
\begin{equation}
\rho_{\infty} = (1-\xi)\frac{1}{2n}\openone_x \otimes \openone_c + \frac{\xi}{2} (\bar{\rho}_+ + \bar{\rho}_-),
\end{equation}  
where 
\begin{eqnarray}
\bar{\rho}_{\pm} &=& \frac{1}{2n} \left(\openone_x \pm \sum_{x=1}^n |x\rangle\langle n-x|\right)\otimes (\openone_c \pm \sigma_y)
\end{eqnarray}
is a product of a position and a coin state.


\subsection{Attractor space and asymptotic dynamics for $\varphi_0 \neq  0$ and $\varphi_1 = 0$}

For $\varphi_0 \neq  0$ and $\varphi_1 = 0$ the eigenvalues of $V$ are $e^{i\varphi_0}$ (non-degenerate) and $1$ ($2n-1$ times degenerate) and the block form of $X_{\lambda}$ is $(2n-1) \oplus 1$, where the one-dimensional space is spanned by $|n\rangle \otimes |0\rangle$. This time it is convenient to use the eigenvectors of the DTQW unitary evolution operator $U$. They are well known, but for the clarity of presentation let us recall their derivation. 

First, it is crucial to note that the evolution operator has translational symmetry. Therefore, its eigenstates are of the form
\begin{equation}
|k_{\pm}\rangle =\left( \frac{1}{\sqrt{n}} \sum_{x=1}^n e^{i\frac{2\pi}{n}xk}|x\rangle \right)\otimes (\alpha_{k_{\pm}}|0\rangle + \beta_{k_{\pm}}|1\rangle),
\end{equation}
where $k=0,1,\ldots,n-1$. As a result, the $2n$-dimensional problem $U|k_{\pm}\rangle = \lambda_{k_{\pm}}|k_{\pm}\rangle$ simplifies to a 2-dimensional one
\begin{equation}\label{2x2ev}
\frac{1}{\sqrt{2}} \begin{pmatrix} e^{i\frac{2\pi}{n}k} & e^{i\frac{2\pi}{n}k} \\ -e^{-i\frac{2\pi}{n}k} & e^{-i\frac{2\pi}{n}k} \end{pmatrix} \begin{pmatrix} \alpha_{k_{\pm}} \\ \beta_{k_{\pm}} \end{pmatrix} =   \lambda_{k_{\pm}} \begin{pmatrix} \alpha_{k_{\pm}} \\ \beta_{k_{\pm}} \end{pmatrix}.
\end{equation}
It has the following solution
\begin{equation}\label{ev}
\lambda_{k_{\pm}} = \frac{1}{\sqrt{2}} \left( \cos\left(\frac{2\pi k}{n}\right) \pm i \sqrt{1+ \sin^2\left(\frac{2\pi k}{n}\right)}\right) \equiv e^{\pm i\phi_{k}},
\end{equation}
where
\begin{equation}
\phi_{k} = \frac{\pi}{2} -\arctan \left(\frac{\cos\left(\frac{2\pi k}{n}\right)}{\sqrt{1+ \sin^2\left(\frac{2\pi k}{n}\right)}}\right).
\end{equation}
Since we assumed that $n$ is odd, note that apart from two eigenvalues $\lambda_{0_{\pm}}$, each of the remaining ones are doubly degenerate. This is because $\lambda_{k_{\pm}}=\lambda_{(n-k)_{\pm}}$, or in other words $\phi_k = \phi_{n-k}$ (where $\phi_0 = \phi_n$). Moreover, the double degeneracy stems from the choice of the coin operator $C$. Finally, the corresponding eigenstates are of the form
\begin{eqnarray}
|k_+\rangle &=&  \left(  \frac{1}{\sqrt{n}} \sum_{x=1}^n e^{i\frac{2\pi}{n}xk}|x\rangle \right)\otimes {\cal N}_k \begin{pmatrix} 1 \\ \chi_k - 1\end{pmatrix}, \\
|k_-\rangle &=& \left(  \frac{1}{\sqrt{n}} \sum_{x=1}^n e^{i\frac{2\pi}{n}xk}|x\rangle \right)\otimes {\cal N}_k \begin{pmatrix}  1-\chi_k^{\ast} \\ 1\end{pmatrix},
\end{eqnarray}
where
\begin{equation}
\chi_k = \sqrt{2} e^{i(\phi_k + \frac{2\pi k}{n})}
\end{equation}
and
\begin{equation}
{\cal N}_k = \frac{1}{\sqrt{\left(4- \chi_k - \chi_k^{\ast}\right)}}.
\end{equation}

Next, we note that due to degeneracy of $|k_{\pm}\rangle$ and $|-k_{\pm}\rangle$ (for $k=1,\ldots,n-1$) one can define the following pairs of states that are also the eigenstates of $U$
\begin{equation}
|\varphi_{k_{\pm}}\rangle = a_{k_{\pm}} |k_{\pm}\rangle +  b_{k_{\pm}} |-k_{\pm}\rangle,
\end{equation}
such that $(\langle n|\otimes\langle 0|) |\varphi_{k_{\pm}}\rangle = 0$. The corresponding coefficients are
\begin{eqnarray}
a_{k_+} &=& \frac{{\cal N}_{-k}}{\sqrt{{\cal N}_{-k}^2+{\cal N}_{k}^2}}, \\
b_{k_+} &=& \frac{{-\cal N}_{k}}{\sqrt{{\cal N}_{-k}^2+{\cal N}_{k}^2}}, \\
a_{k_-} &=& \frac{{\cal N}_{-k}(1-\chi_{-k}^{\ast})}{\sqrt{{\cal N}_{-k}^2|1-\chi_{-k}^{\ast}|^2+{\cal N}_{k}^2|1-\chi_{k}^{\ast}|^2}}, \\
b_{k_-} &=& \frac{{-\cal N}_{k}(1-\chi_{k}^{\ast})}{\sqrt{{\cal N}_{-k}^2|1-\chi_{-k}^{\ast}|^2+{\cal N}_{k}^2|1-\chi_{k}^{\ast}|^2}}.
\end{eqnarray}
If one applied $V$ to these states, the action would be the same as if one applied $\openone_x\otimes\openone_c$. Therefore, the states $|\varphi_{k_{\pm}}\rangle$ are eigenstates of $U_0$ and $U_1$. Moreover, $U_0|\varphi_{k_{\pm}}\rangle=U_1|\varphi_{k_{\pm}}\rangle = \lambda_{k_{\pm}}|\varphi_{k_{\pm}}\rangle$, hence the random unitary channel (\ref{ruc}) generates a unitary evolution of such states.

The above results lead to immediate conclusion that the attractor space is a $(n-1)$-dimensional Hilbert subspace spanned by the vectors $\{|\varphi_{k_{\pm}}\rangle\}_{k=1}^{(n-1)/2}$. Apart from $X_{\lambda = 1}^{(1)} = \openone_x \otimes \openone_c$, the attractor space consists of the following operators
\begin{equation}
X_{k_{\pm},k'_{\pm}} = |\varphi_{k_{\pm}}\rangle\langle\varphi_{k'_{\pm}}|,
\end{equation}
and the corresponding eigenvalues are $\lambda_{k_{\pm},k'_{\pm}}=\lambda_{k_{\pm}}\lambda_{k'_{\pm}}^{\ast}$. Moreover, note that $X_{k_{\pm},k'_{\pm}}^{\dagger}$ also belongs to the attractor space and the corresponding eigenvalue is $\lambda_{k_{\pm},k'_{\pm}}^{\ast}$. Therefore, for an initial state $\rho_0$, such that ${\text Tr}\{X_{k_{\pm},k'_{\pm}} \rho_0\} = {\text Tr}\{X^{\dagger}_{k_{\pm},k'_{\pm}} \rho_0\}^{\ast}\neq 0$, one can observe oscillatory asymptotic dynamics with a period determined by $\lambda_{k_{\pm},k'_{\pm}}$. In addition, any operator of the following form
\begin{equation}
X_{\lambda=1} = \sum_{k=1}^{(n-1)/2}\sum_{j=\pm}\gamma_{k_j}|\varphi_{k_{j}}\rangle\langle\varphi_{k_{j}}|
\end{equation}
is in the attractor space and corresponds to the eigenvalue $\lambda = 1$.

Note that each eigenvector $|\varphi_{k_{\pm}}\rangle$ corresponds to a different eigenvalue. For $n=3$ there are only two such vectors, therefore the reduced asymptotic dynamics of the coin, presented in Fig. \ref{fig3}, is relatively simple since it consists of only one oscillatory term. On the other hand, in higher dimensions the asymptotic dynamics consists of many oscillations. In addition, due to discreteness of dynamics and the irrationality of the eigenvalues $\lambda_{k_{\pm},k'_{\pm}}$, the oscillatory patterns may take complex forms (see Fig. \ref{fig3}).

Finally, let us show that states $|\varphi_{k_{\pm}}\rangle$ are entangled. They are pure, therefore the entanglement can be verified by calculating the purity of the reduced density matrix of the coin degree of freedom
\begin{equation}
{\text Tr}_x \{|\varphi_{k_+}\rangle\langle \varphi_{k+}|\} = {\cal N}\begin{pmatrix}  1 & c^{\ast} \\ c & b \end{pmatrix},
\end{equation}
where
\begin{eqnarray}
c &=& \frac{\chi_k+\chi_{-k}}{2} - 1, \\
b &=& 1-c-c^{\ast},
\end{eqnarray}
and ${\cal N}=1/(1+b)$. Similarly,
\begin{equation}
{\text Tr}_x \{|\varphi_{k_-}\rangle\langle \varphi_{k_-}|\} = {\cal N}\begin{pmatrix}  b & c^{\ast} \\ c & 1 \end{pmatrix}.
\end{equation}
The purity of both reduced density matrices is
\begin{equation}
P= \frac{1 + 2|c|^2 + b^2}{1 + 2b + b^2},
\end{equation}
therefore the state is entangled if $|c|^2 < b$ (determinant of both matrices is greater than zero). Simple substitution shows that the above is equivalent to
\begin{equation}
\cos^2\left(\frac{2\pi k}{n}\right) < 1,
\end{equation}
which is always true, since we consider $k=1,\ldots,n-1$ (recall that $n$ is odd).


\section{Example: 3-cycle}

\begin{figure*}[t]
\includegraphics[scale=0.35]{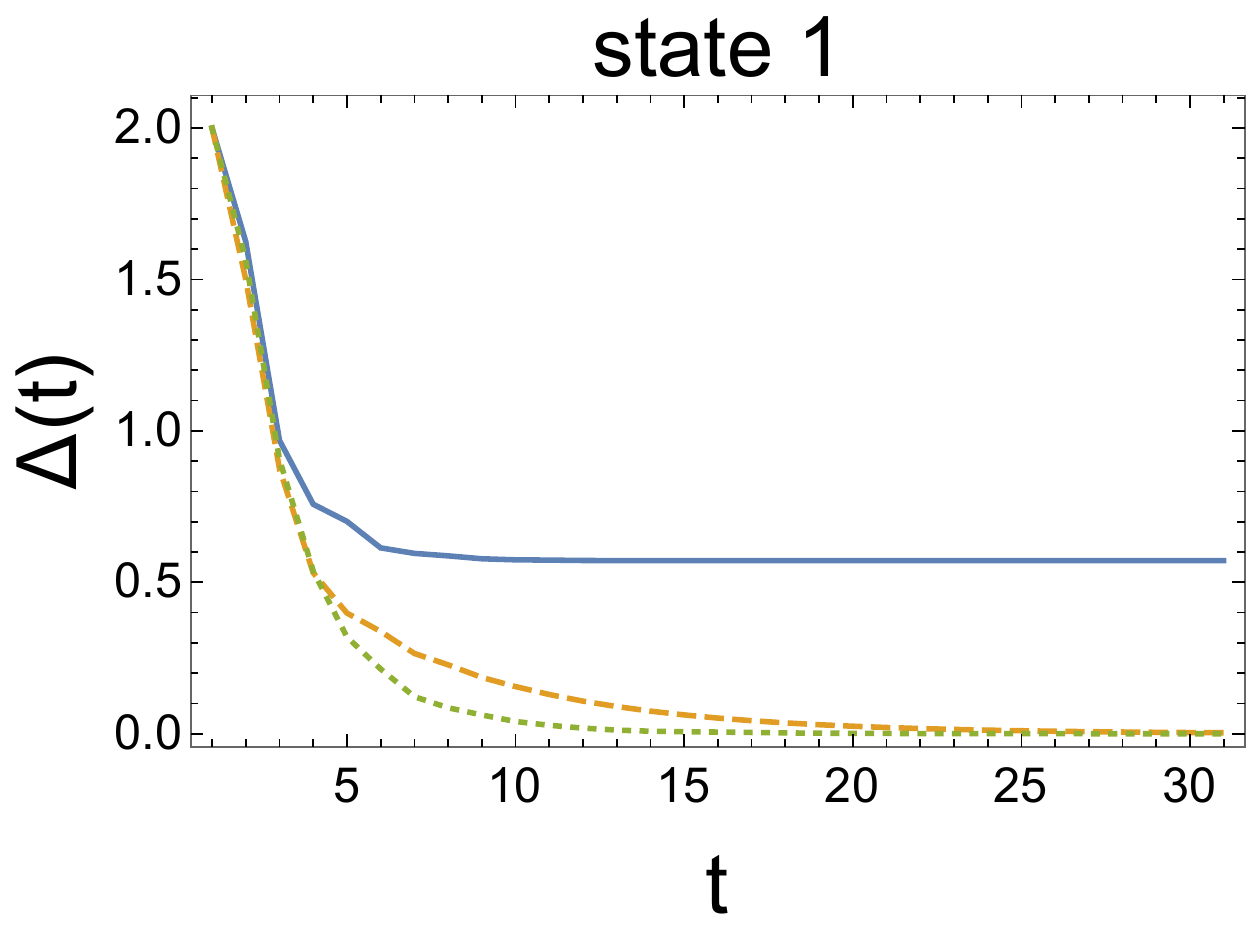}\includegraphics[scale=0.35]{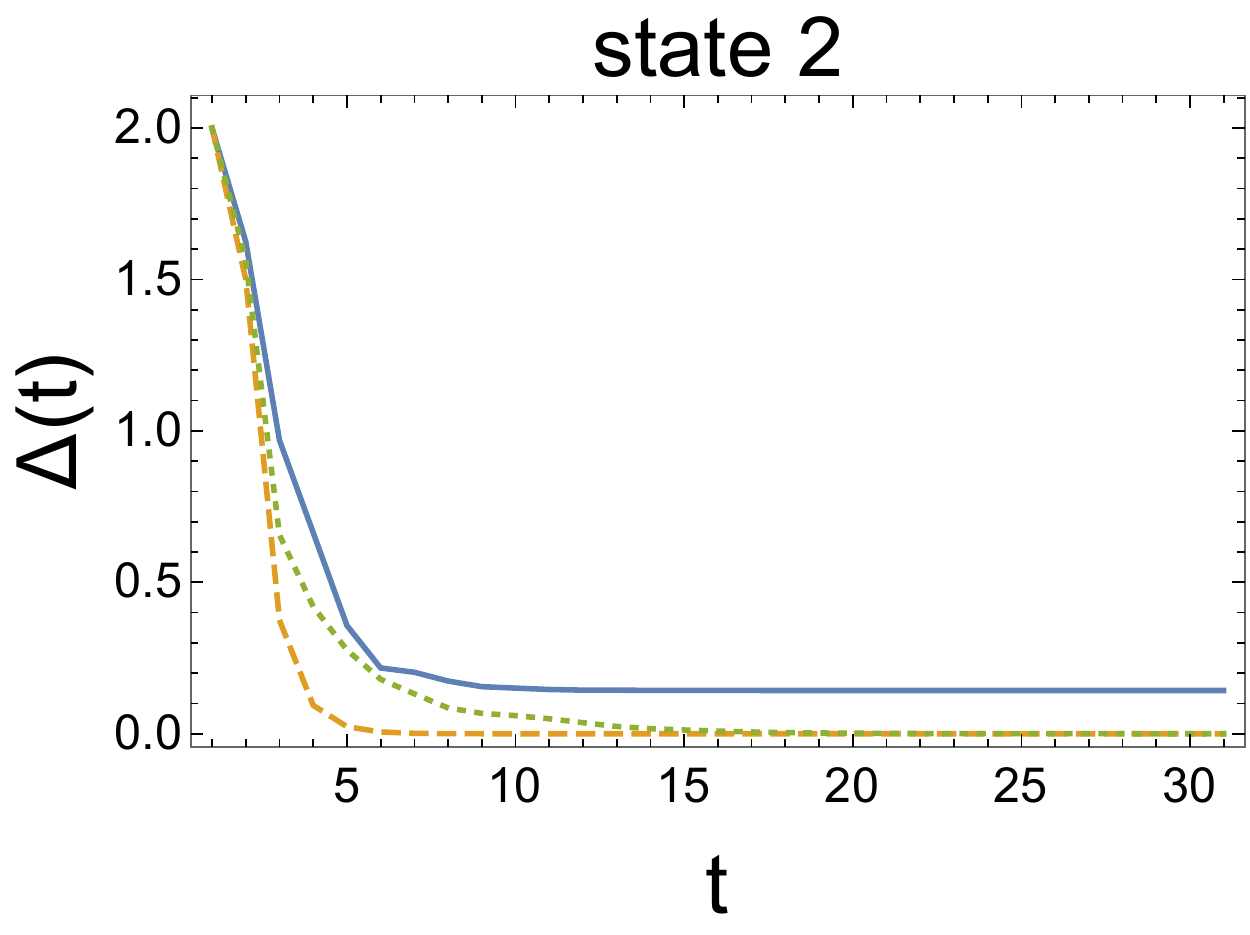}\includegraphics[scale=0.35]{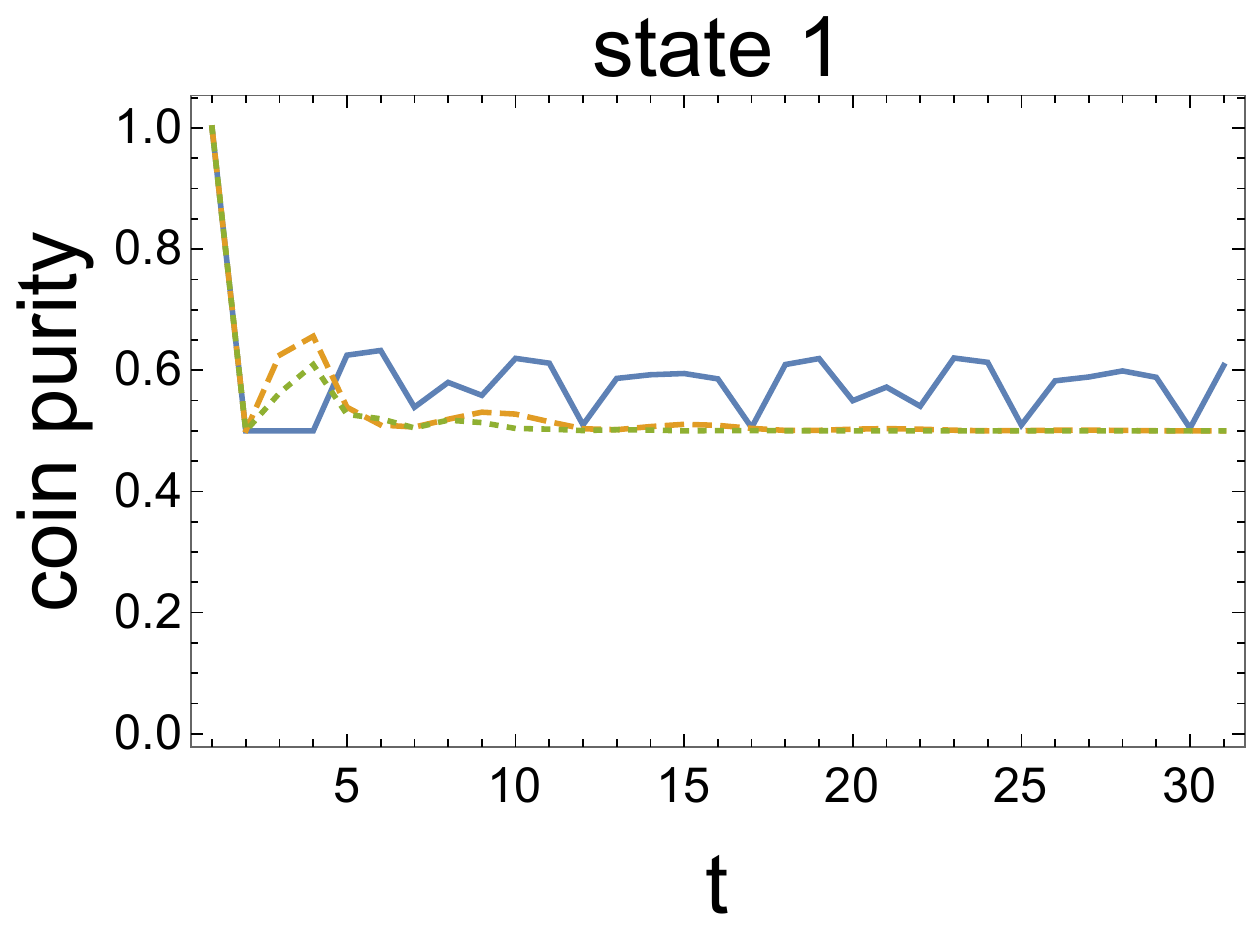}\includegraphics[scale=0.35]{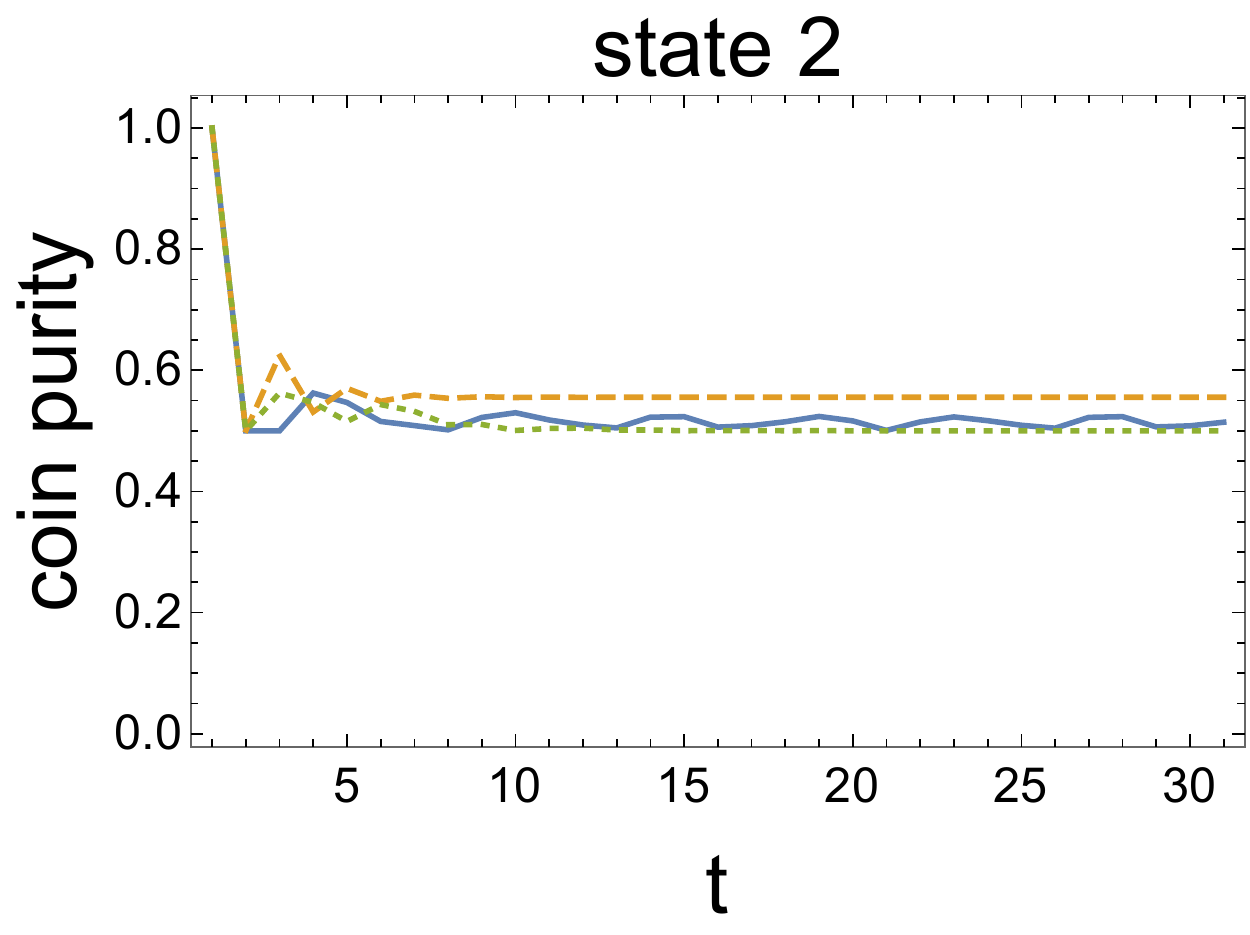}
\caption{The evolution of $\Delta(t)$ and the purity of the coin density matrix for the walk on 3-cycle. The time is discrete, however for the purpose of presentation the points were connected. Solid blue line corresponds to the case $\varphi_1 = 0$, green dotted to $\varphi_1 = \pi/2$ and orange dashed to $\varphi_1 = \pi$. The walk starts in the state $|\psi_0\rangle=|3\rangle\otimes|c_0\rangle$. Two initial coin states are considered: $|c_0\rangle=|1\rangle$ (state 1) and $|c_0\rangle=\frac{1}{\sqrt{2}}(|0\rangle + i|1\rangle)$ (state 2). \label{fig4}}
\end{figure*}

In this section we discuss in more details the simplest case -- the asymptotic dynamics on 3-cycle. Although the model is described by three parameters, to observe all possible behaviors one can fix $\eta$ and one phase. Therefore, from now on we fix $\eta = 1/2$, $\varphi_0 = \pi$ and vary $\varphi_1 \in [0,\pi]$. The same can be done for general n-cycles. If $\varphi_1 = 0$ one can observe oscillatory asymptotic dynamics, if $\varphi_1 = \varphi_0$ one can observe relaxation to a non-maximally mixed stationary state and if $0 < \phi_1 < \phi_0$ one observes relaxation to a maximally mixed state.

For simplicity we assume that the initial state is $|\psi_0\rangle=|3\rangle\otimes |c_0\rangle$, i.e., it is localized at position $x=3$ and the initial state of the coin $|c_0\rangle$ is pure. In order to visualize the system's tendency to the asymptotic dynamics we consider the following property
\begin{equation}
\Delta(t)={\text Tr}\{\left(\rho(t+1)-\rho(t)\right)^2\},
\end{equation} 
which measures how close the two consecutive states are. In addition, we consider the purity of the reduced density matrix of the coin. In Fig. \ref{fig4} we show the evolution of $\Delta(t)$ and the purity of the reduced density matrix of the coin for two different initial coin states. We see that the system reaches the equilibrium after less than 20 steps.


\subsection{Asymptotic dynamics for $\varphi_1 \neq  0$}

If $0 < \phi_1 < \phi_0$ the asymptotic behavior is state independent -- there is just one fixed point corresponding to $\frac{1}{6}\openone_x \otimes \openone_c$. The case $\varphi_1=\varphi_0=\pi$ is more interesting, because the asymptotic behavior strongly depends on the initial state. The asymptotic state, given by Eq. (\ref{asymptotic}), is 
\begin{equation}
\rho_{\infty}=\frac{1}{6}\left(\openone_x\otimes\openone_c + \xi (|1\rangle\langle 2| + |2\rangle\langle 1| + |3\rangle\langle 3|) \otimes \sigma_y \right).
\end{equation}
The above is a mixed state with eigenvalues $(1\pm \xi)/6$, each triply degenerate. In particular, the overlap $-1 \leq \xi \leq 1$ is in this case given by $\xi=\langle c_0|\sigma_y|c_0\rangle$. For example, if $|c_0\rangle = |1\rangle$ the overlap is $\xi=0$, whereas for $|c_0\rangle = \frac{1}{\sqrt{2}}(|0\rangle + i|1\rangle)$ the overlap is $\xi=1$. The corresponding asymptotic reduced state of the coin is
\begin{equation}
\rho_{c_{\infty}}=\frac{1}{2}\openone_c + \frac{\xi}{6}\sigma_y,
\end{equation}
which confirms our fitting from Eq. (\ref{fitting}). 


\subsection{Asymptotic dynamics for $\varphi_1 = 0$}

For $\varphi_1 = 0$ the asymptotic dynamics is oscillatory, which is presented in Fig. {\ref{fig4}}. The attractor space is two-dimensional and is spanned by
\begin{eqnarray}
|\varphi_{1_{+}}\rangle =& \frac{1}{\sqrt{7}} & \left(|1\rangle\otimes |0\rangle - \frac{1+i\sqrt{7}}{2}|1\rangle\otimes |1\rangle - |2\rangle\otimes |0\rangle \right. \nonumber \\
&+& \left. |2\rangle\otimes |1\rangle - \frac{1-i\sqrt{7}}{2}|3\rangle\otimes |1\rangle\right)  \label{phi+}, \\ 
|\varphi_{1_{-}}\rangle =&  \frac{1}{\sqrt{7}} & \left(|1\rangle\otimes |0\rangle - \frac{1-i\sqrt{7}}{2}|1\rangle\otimes |1\rangle - |2\rangle\otimes |0\rangle \right. \nonumber \\
&+& \left. |2\rangle\otimes |1\rangle - \frac{1+i\sqrt{7}}{2}|3\rangle\otimes |1\rangle  \right)  \label{phi-},
\end{eqnarray}
corresponding to the eigenvalues $\lambda_{1_{\pm}} = \frac{1}{2\sqrt{2}}(-1\pm i\sqrt{7})$. Notice that both vectors are orthogonal to $|3\rangle\otimes|0\rangle$.

We define the following operators, that are the eigenvectors of the random unitary evolution,
\begin{eqnarray}
\Pi_{+} &=& |\varphi_{1_+}\rangle\langle \varphi_{1_+}|, \\
\Pi_{-} &=& |\varphi_{1_-}\rangle\langle \varphi_{1_-}|, \\
X_{+} &=& |\varphi_{1_+}\rangle\langle \varphi_{1_-}|, \\
X_{-} &=& X_{+}^{\dagger}.
\end{eqnarray}
These operators have the following overlap with the initial state $|\psi_0\rangle$
\begin{eqnarray}
p_+ &=& \langle \psi_0|\Pi_{+} |\psi_0\rangle, \\
p_- &=& \langle \psi_0|\Pi_{-} |\psi_0\rangle, \\
\kappa &=& \langle \psi_0|X_{+} |\psi_0\rangle.
\end{eqnarray}
In addition, we define
\begin{eqnarray}
\bar{I} = \openone_x\otimes\openone_c - |\varphi_{1_+}\rangle\langle \varphi_{1_+}| - |\varphi_{1_-}\rangle\langle \varphi_{1_-}|.
\end{eqnarray}
The asymptotic state is given by
\begin{eqnarray}
\rho_{\infty}(t) &=& \frac{1-p_+ -p_-}{4}\bar{I} + p_{+}\Pi_{+} + p_{-}\Pi_{-} \nonumber \\
&+& \kappa \Lambda^t X_{+} + \kappa^{\ast}\Lambda^{- t}X_{-},
\end{eqnarray}
where $\Lambda = \lambda_{1_+}\lambda_{1_-}^{\ast}=\lambda_{1_+}^2$. For the initial coin state $|c_0\rangle = |1\rangle$ the parameters determining the asymptotic state are $p_+ = p_- = 2/7$ and $\kappa = -\frac{1}{14}(3+i\sqrt{7})$. In general, if the initial coin state were $|c_0\rangle = \alpha|0\rangle + \beta |1\rangle$, the above parameters would be multiplied by $|\beta|^2$. 

\begin{figure}[t]
\includegraphics[scale=0.55]{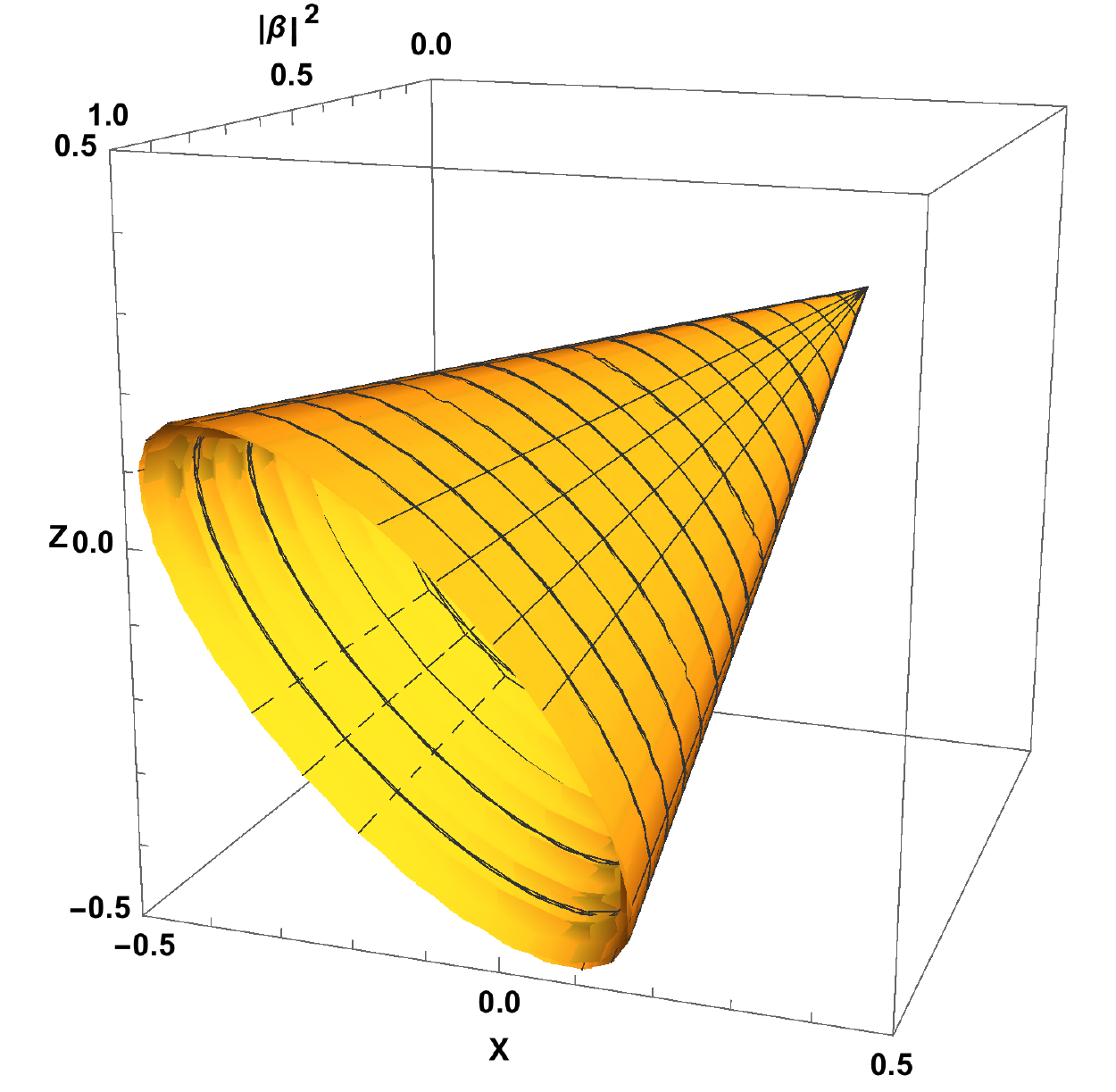}
\caption{Bloch sphere representation of the attractor for the coin subsystem. The walk is on 3-cycle, $\varphi_1=0$ and the initial state is $|\psi_0\rangle = |3\rangle\otimes(\alpha|0\rangle + \beta|1\rangle)$. The plot represents the asymptotic trajectory of the corresponding Bloch vector in the XZ-plane as a function of $|\beta|^2$. \label{fig5}}
\end{figure}

Let us also discuss the asymptotic dynamics of the coin subsystem. Straightforward calculations show that for the initial coin state $|c_0\rangle = \alpha|0\rangle + \beta |1\rangle$ one gets
\begin{eqnarray}
{\text Tr}\{\rho_{c_{\infty}}(t)\sigma_x\} &=&  \frac{21-36|\beta|^2+4|\beta|^2{\text Re}(\omega \Lambda^t)}{98},\\
{\text Tr}\{\rho_{c_{\infty}}(t)\sigma_y\} &=& 0, \\
{\text Tr}\{\rho_{c_{\infty}}(t)\sigma_z\} &=&  \frac{21-36|\beta|^2+32|\beta|^2 {\text Re}(\Lambda^{t+1})}{98},
\end{eqnarray}
where $\omega = 1+3i\sqrt{7}$. Therefore, the evolution takes place in the XZ-plane of the Bloch sphere and the path of the corresponding Bloch vector is ellipsoidal (see Fig. \ref{fig5}), which confirms our previous numerical simulations. 

\begin{figure}[t]
\includegraphics[scale=0.55]{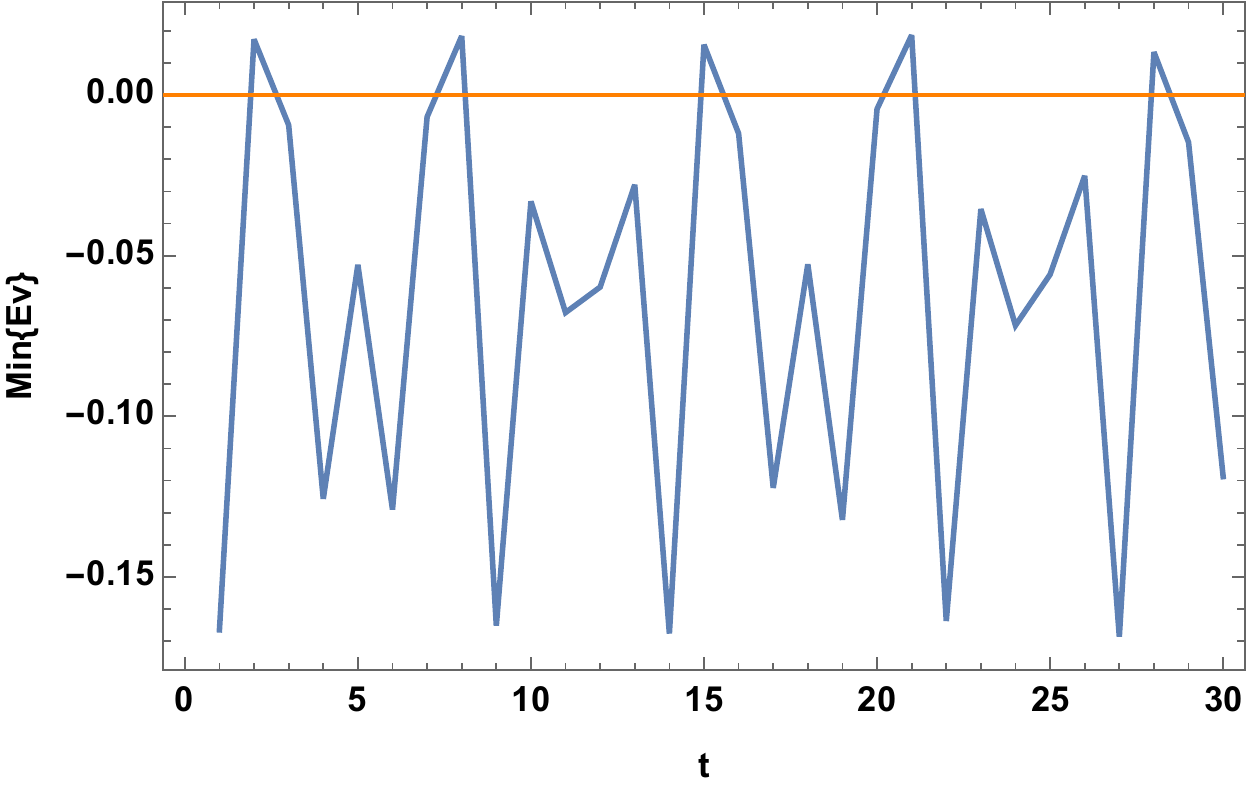}
\caption{Time dependence of the smallest eigenvalue of the partially transposed state  $\rho_{\infty}^{\Gamma}(t)$. The initial state is $|\psi_0\rangle = |3\rangle\otimes|1\rangle$ and the walk is on 3-cycle with $\varphi_1 = 0$. The points were connected for a better visualization. Negative value implies that the state $\rho_{\infty}(t)$ is entangled. \label{fig6}}
\end{figure}

Finally, let us discuss the position-coin entanglement in the asymptotic state. Since for a 3-cycle the system is made of a qubit (coin) and a qutrit (position), the entanglement can be verified via detection of a negative eigenvalue of the partially transposed state $\rho_{\infty}^{\Gamma}(t)$ \cite{Horo}. In Fig. \ref{fig6} we plotted how the smallest eigenvalue of  $\rho_{\infty}^{\Gamma}(t)$ changes in time for the initial state $|\psi_0\rangle = |3\rangle\otimes|1\rangle$. Such an initial state guarantees the highest overlap with the attractor space for a system that is in a product state and is initially localized at a single position. We see that for most time steps the smallest eigenvalue is negative, therefore the system is entangled. In particular, for 30 time steps there are only five cases in which the system is not entangled, which confirms that the asymptotic behavior can be non-classical.

\section{Conclusions and outlook}

We showed that a simple modification of a unitary DTQW evolution to a non-unitary open dynamics can lead to a collection of diverse asymptotic dynamical behaviors. These behaviors depend on the parameters of the evolution. For majority of parameters the dynamics tends to a maximally mixed state, which is the same as the stationary state of the classical random walk on a n-cycle. However, there are some sets of parameters for which the system exhibits a non-classical asymptotic behavior. In particular, we showed that the system can fall onto an attractive orbit that is made mostly of entangled states. Therefore, the model proposed by us can be used to generate and sustain entanglement via an open system evolution. 

The attractive orbit can be thought of as a limit cycle that is well visible in the Bloch vector representation of the coin degree of freedom. Interestingly, the coin is a two-level system, a qubit, which was recently considered in the context of quantum synchronization \cite{Sync}. More precisely, the existence of a limit cycle is a prerequisite for synchronization and the authors of \cite{Sync} argued that qubit dynamics in the presence of gains and losses does not have limit cycles. However, our result seem to contradict this statement. 

There are two reasons for the apparent contradiction. Firstly, the attractive orbit is not a limit cycle per se because it depends not only on the parameters of the evolution, but also on the initial state. For a different initial state the system may end up on some other closed orbit in the vicinity of the previous one. This is in contrast to the standard definition of the limit cycle which is supposed to be an isolated closed orbit independent of an initial state \cite{Strogatz}.  Secondly, in our case the total (Markovian) dynamics describes the evolution of a qubit and an auxiliary position degree of freedom, therefore the alleged limit cycle is not bound to the qubit subspace. Still, the reduced dynamics of the qubit seems to posses such cycle, however in this case the evolution is non-Markovian. Therefore, it would be interesting to investigate how non-Markovian evolutions can be used for quantum synchronization. 


\section*{Acknowledgements}

This work is supported by the Ministry of Science and Higher Education in Poland (science funding scheme 2016-2017 project no. 0415/IP3/2016/74).



\end{document}